\documentclass[12pt]{article}
\usepackage[dvips]{color}
\usepackage{epsfig}
\usepackage{amsmath}
\usepackage{graphicx}
\def\Box{\hbox{$\rlap{$\sqcup$}\sqcap$}}
\begin{document}
\title{\bf Formulation of Electrodynamics with an External Source in the Presence of a Minimal Measurable Length}

\author{S. K. Moayedi $^{a}$\thanks{E-mail: s-moayedi@araku.ac.ir}\hspace{1mm}
, M. R. Setare $^{b}$ \thanks{E-mail:
rezakord@ipm.ir}\hspace{1mm},
 B. Khosropour $^{a}$\thanks{E-mail: b-khosropour@phd.araku.ac.ir}\hspace{1.5mm}  \\
$^a$ {\small {\em  Department of Physics, Faculty of Sciences,
Arak University, Arak 38156-8-8349, Iran}}\\
$^{b}${\small {\em Department of Science, Campus of Bijar,
University of Kurdistan Bijar, Iran
}}\\
}
\date{\small{}}
\maketitle
\begin{abstract}
\noindent In a series of papers, Quesne  and Tkachuk (J. Phys. A:
Math. Gen. \textbf{39}, 10909 (2006); Czech. J. Phys. \textbf{56},
1269 (2006)) presented a $D+1$-dimensional
$(\beta,\beta')$-two-parameter Lorentz-covariant deformed algebra
which leads to a nonzero minimal measurable length. In this paper,
the Lagrangian formulation of electrodynamics in a
$3+1$-dimensional space-time described by Quesne-Tkachuk algebra
is studied in the special case $\beta'=2\beta$ up to first order
over the deformation parameter $\beta$. It is demonstrated that
at the classical level there is a similarity between
electrodynamics in the presence of a minimal measurable length
(generalized electrodynamics) and Lee-Wick electrodynamics. We
obtain the free space solutions of the inhomogeneous Maxwell's
equations in the presence of a minimal length. These solutions
describe two vector particles (a massless vector particle and a
massive vector particle). We estimate two different upper bounds
on the isotropic minimal length. The first upper bound is near to
the electroweak length scale $(\ell_{electroweak}\sim 10^{-18}\,
m)$, while the second one is near to the length scale for the
strong interactions $(\ell_{strong}\sim 10^{-15}\, m)$. The
relationship between the Gaete-Spallucci nonlocal electrodynamics
(J. Phys. A: Math. Theor. \textbf{45}, 065401 (2012)) and
electrodynamics with a minimal length is investigated.
\\

\noindent
\hspace{0.35cm}



\end{abstract}

\section{Introduction}
The unification between general theory of relativity and the
Standard Model of particle physics is one of the most important
problems in theoretical physics [1]. This unification predicts the
existence of a minimal measurable length on the order of the
Planck length. Also, recent studies in perturbative string theory
and loop quantum gravity suggest that there is a minimal length
scale in nature [2].\\ Today's we know that the existence of a
minimal measurable length leads to an extended uncertainty
principle. This extended uncertainty principle can be written as
\begin{equation}
\triangle X \triangle
P\geq\frac{\hbar}{2}\left[1+a_{1}(\frac{l_{P}}{\hbar})^{2}(\Delta
P)^{2}+a_{2}(\frac{l_{P}}{\hbar})^{4}(\Delta P)^{4}+\cdots\right],
\end{equation}
where $l_{P}$ is the Planck length and $a_{i}\;,\forall i\in
\{1,2, \cdots \}$ are positive numerical constants [3,4]. If we
keep only the first two terms on the right hand-side of (1), we
will obtain the usual generalized uncertainty principle (GUP) as
follows:
\begin{equation}
\triangle X \triangle
P\geq\frac{\hbar}{2}\left[1+a_{1}(\frac{l_{P}}{\hbar})^{2}(\Delta
P)^{2}\right].
\end{equation}
It is obvious that in (2), $\triangle X$ is always greater than
$(\triangle X)_{min}=\sqrt{a_{1}}\;l_{P}$. Many physicists believe
that reformulation of quantum field theory in the presence of a
minimal measurable length leads to a divergence free quantum field
theory [5-7]. In the recent years, reformulation of quantum
mechanics, gravity and quantum field theory in the presence of a
minimal measurable length have been studied extensively [4-18].
The first attempt to construct the electromagnetic field in
quantized space-time was made by H. S. Snyder [19]. In a previous
work [14] we studied formulation of an electrostatic field with a
charge density in the presence of a minimal length based on the
Kempf algebra. In the present work we study formulation of
electrodynamics with an external source in the presence of a
minimal measurable length based on the Quesne-Tkachuk algebra.
The organization of our paper is as follows. In Sec. 2, the
$D+1$-dimensional $(\beta,\beta')$-two-parameter
Lorentz-covariant deformed algebra introduced by Quesne and
Tkachuk is studied and it is shown that the Quesne-Tkachuk
algebra leads to a minimal measurable length [20,21]. In Sec. 3,
the Lagrangian formulation of electrodynamics with an external
source in a $3+1$-dimensional space-time described by
Quesne-Tkachuk algebra is introduced in the special case
$\beta'=2\beta$, in which the position operators commute to first
order in $\beta$. We show that at the classical level there is a
similarity between electrodynamics in the presence of a minimal
measurable length and Lee-Wick electrodynamics. In Sec. 4, the
free space solutions of the inhomogeneous Maxwell's equations in
the presence of a minimal measurable length are obtained. These
solutions describe two different particles, a massless vector
particle and a massive vector particle. In Sec. 5, we obtain two
different upper bounds on the isotropic minimal length. One of
these upper bounds on the isotropic minimal length is near to the
electroweak length scale $(\ell_{electroweak}\sim 10^{-18}\, m)$.
The second upper bound on the isotropic minimal length is near to
the length scale for the strong interactions $(\ell_{strong}\sim
10^{-15}\, m)$. In Sec. 6, we investigate the relation between
electrodynamics in the presence of a minimal measurable length
and the concept of nonlocality in electrodynamics. Our
conclusions are presented in Sec. 7. SI units are used throughout
this paper.

\section{Lorentz-Covariant Deformed Algebra with a Minimal Observable Distance}
Recently, Quesne and Tkachuk have introduced a Lorentz-covariant
deformed algebra which describes a $D+1$-dimensional quantized
space-time [20,21]. The Quesne-Tkachuk algebra in a
$D+1$-dimensional space-time is specified by the following
generalized commutation relations:
\begin{eqnarray}
\left[X^{\mu},P^{\nu}\right] &=&-i\hbar[(1-\beta P_{\rho}P^{\rho})g^{\mu\nu}-\beta'P^{\mu}P^{\nu}], \\
\left[X^{\mu},X^{\nu}\right] &=& i\hbar\
\frac{2\beta-\beta'-(2\beta+\beta')\beta P_{\rho}P^{\rho}
}{1-\beta P_{\rho}P^{\rho}}(P^{\mu}X^{\nu}-P^{\nu}X^{\mu}), \\
\left[P^{\mu},P^{\nu}\right] &=& 0,
\end{eqnarray}

where $\mu,\; \nu, \; \rho=0,1,2,\cdots,D$ and $\beta ,\; \beta'$
are two non-negative deformation parameters $(\beta,\beta' \geq 0)$.
In the above equations $\beta$ and $\beta'$ are constant parameters
with dimension $(momentum)^{-2}$. Also, $X^{\mu}$ and $P^{\mu}$ are
position and momentum operators in the deformed space and
$g_{\mu\nu}=g^{\mu\nu}=diag(1,-1,-1,\cdots,-1)$. In the special case
where $D=3$ and $\beta=0$, the Quesne-Tkachuk algebra (3)-(5)
reduces to the Snyder algebra [22].\\ An immediate consequence of
relation (3) is the appearance of an isotropic minimal length which
is given by
\begin{equation}
(\triangle X^{i})_{0}=\hbar\sqrt{(D\beta+\beta')
[1-\beta\langle(P^{0})^{2}\rangle]}\quad , \quad\forall i\in
\{1,2, \cdots ,D\}.
\end{equation}
In Ref. [23], Tkachuk introduced a representation which satisfies
the generalized commutation relations (3)-(5) up to first order
in deformation parameters $\beta$ and $\beta'$.\\
The Tkachuk representation is given by
\begin{eqnarray}
X^{\mu} &=& x^{\mu}-
\frac{2\beta-\beta'}{4}(x^{\mu} p_{\rho}p^{\rho}+p_{\rho}p^{\rho}x^{\mu}), \\
P^{\mu} &=& (1-\frac{\beta'}{2}p_{\rho}p^{\rho})p^{\mu},
\end{eqnarray}
where $x^{\mu}$ and $p^{\mu}=i\hbar \frac{\partial}{\partial
x_{\mu}}=i\hbar\partial^{\mu} $ are position and momentum
operators in ordinary relativistic quantum mechanics. In this
study, we consider the special case $\beta'=2\beta$, in which the
position operators commute to first order in deformation
parameter $\beta$, i.e., $[X^{\mu},X^{\nu}]=0$. In this linear
approximation, the Quesne-Tkachuk algebra becomes
\begin{eqnarray}
\left[X^{\mu},P^{\nu}\right] &=&-i\hbar[(1-\beta P_{\rho}P^{\rho})g^{\mu\nu}-2\beta P^{\mu}P^{\nu}], \\
\left[X^{\mu},X^{\nu}\right] &=& 0, \\
\left[P^{\mu},P^{\nu}\right] &=& 0.
\end{eqnarray}
The following representations satisfy (9)-(11), in the first order
in $\beta$:
\begin{eqnarray}
X^{\mu} &=& x^{\mu}, \\
P^{\mu} &=& (1-\beta p_{\rho}p^{\rho})p^{\mu}.
\end{eqnarray}
Note that the representations (7), (8) and (12), (13) coincide
when $\beta'=2\beta$.

\section{Lagrangian Formulation of Electrodynamics with an External Source in the Presence of a Minimal Length Based on
the Quesne-Tkachuk Algebra} The Lagrangian density for a massless
vector field $A^{\mu}=(\frac{1}{c}\phi,\textbf{A})$ with an external
source $J^{\mu}=(c\rho,\textbf{J})$ in a $3+1$-dimensional
space-time is [24]
\begin{equation}
{\cal L}=-\frac{1}{4\mu_{0}}F_{\mu\nu}F^{\mu\nu}-J^{\mu}A_{\mu},
\end{equation}
where $F_{\mu\nu}=\partial_{\mu}A_{\nu}-\partial_{\nu}A_{\mu}$ is
the electromagnetic field tensor. In a $3+1$-dimensional
space-time the components of the electromagnetic field tensor
$F_{\mu\nu}$ can be written as
\begin{eqnarray}
F_{\mu\nu}&=&
\left( {\begin{array}{cccc}
   0 & {E_{x}}/{c}\ & {E_{y}}/{c}  & {E_{z}}/{c}  \\
   -{E_{x}}/{c} & 0 & -B_{z} &B _{y} \\
   -{E_{y}}/{c} & B_{z} & 0 & -B_{x} \\
   -{E_{z}}/{c} & -B_{y} & B_{x} & 0
    \end{array}} \right).
\end{eqnarray}
The Euler-Lagrange equation for the vector field $A^{\mu}$ is
\begin{equation}
\frac{\partial{\cal L}}{\partial A_{\lambda}
}-\partial_{\rho}\left(\frac{\partial{\cal
L}}{\partial(\partial_{\rho}A_{\lambda})}\right)=0.
\end{equation}
If we substitute the Lagrangian density (14) in the
Euler-Lagrange equation (16), we will obtain the inhomogeneous
Maxwell's equations as follows:
\begin{equation}
\partial_{\rho}F^{\rho\lambda}=\mu_{0}J^{\lambda}.
\end{equation}
The electromagnetic field tensor $F_{\mu\nu}$ satisfies the
Bianchi identity
\begin{equation}
\partial_{\mu}F_{\nu\lambda}+\partial_{\nu}F_{\lambda\mu}+\partial_{\lambda}F_{\mu\nu}=0.
\end{equation}
Equation (18) leads to the homogeneous Maxwell's equations. Now,
we obtain the Lagrangian density for electrodynamics in the
presence of a minimal observable distance based on the
Quesne-Tkachuk algebra. For this purpose, let us write the
Lagrangian density (14) by using the representations (12) and
(13), i.e.,
\begin{eqnarray}
x^{\mu}\longrightarrow  X^{\mu}&=&x^{\mu}, \\
\partial^{\mu}\longrightarrow\nabla^{\mu}&:=&(1+\beta\hbar^{2}\Box)\partial^{\mu},
\end{eqnarray}
where $\Box:=\partial_{\mu}\partial^{\mu}$ is the d'Alembertian
operator. The result reads
\begin{eqnarray}
{\cal L} &=&
-\;\frac{1}{4\mu_{0}}\;(\nabla_{\mu}A_{\nu}-\nabla_{\nu}A_{\mu})(\nabla^{\mu}A^{\nu}-\nabla^{\nu}A^{\mu})-J^{\mu}A_{\mu}
\nonumber
\\ &=&- \;\frac{1}{4\mu_{0}}\;[(1+\beta
\hbar^{2}\Box)\partial_{\mu}A_{\nu}-(1+\beta\hbar^{2}\Box)\partial_{\nu}A_{\mu}]
\nonumber \\
 & &
[(1+\beta\hbar^{2}\Box)\partial^{\mu}A^{\nu}-(1+\beta\hbar^{2}\Box)\partial^{\nu}A^{\mu}]-J^{\mu}A_{\mu}
\nonumber \\
 &=&-\; \frac{1}{4\mu_{0}}F_{\mu\nu}F^{\mu\nu}- \frac{1}{4\mu_{0}}(\hbar\sqrt{2\beta})^{2}F_{\mu\nu}\;\Box F^{\mu\nu}
\nonumber \\
 & & -J^{\mu}A_{\mu}+{\cal O}\left((\hbar\sqrt{2\beta})^{4}\right).
\end{eqnarray}
The term
$-\frac{1}{4\mu_{0}}(\hbar\sqrt{2\beta})^{2}F_{\mu\nu}\;\Box
F^{\mu\nu} $ in the above Lagrangian can be considered as a
minimal length effect.\\ If we neglect terms of order
$(\hbar\sqrt{2\beta})^{4}$ and higher in (21), we will obtain the
following Lagrangian density
\begin{equation}
{\cal L}=-\;\frac{1}{4\mu_{0}}
F_{\mu\nu}F^{\mu\nu}-\frac{1}{4\mu_{0}}(\hbar\sqrt{2\beta})^{2}F_{\mu\nu}\;\Box
F^{\mu\nu}-J^{\mu}A_{\mu}.
\end{equation}
The Lagrangian density (22) is similar to the Abelian Lee-Wick model
which was introduced by Lee and Wick as a finite theory of quantum
electrodynamics [25-29]. The equation (22) can be written as
follows:
\begin{equation}
{\cal L}=-\;\frac{1}{4\mu_{0}}
F_{\mu\nu}F^{\mu\nu}+\frac{1}{4\mu_{0}}(\hbar\sqrt{2\beta})^{2}(\partial_{\alpha}F_{\mu\nu})
(\partial^{\alpha}F^{\mu\nu})+\partial_{\alpha}\chi^{\alpha}-J^{\mu}A_{\mu},
\end{equation}
where
\begin{equation}
\chi^{\alpha}:=-\;\frac{1}{4\mu_{0}}(\hbar\sqrt{2\beta})^{2}F_{\mu\nu}\;\partial^{\alpha}
F^{\mu\nu}.
\end{equation}
After dropping the total derivative term
$\partial_{\alpha}\chi^{\alpha}$, the Lagrangian density (23) will
be equivalent to the following Lagrangian density:
\begin{equation}
{\cal L}=-\;\frac{1}{4\mu_{0}}
F_{\mu\nu}F^{\mu\nu}+\frac{1}{4\mu_{0}}(\hbar\sqrt{2\beta})^{2}(\partial_{\alpha}F_{\mu\nu})
(\partial^{\alpha}F^{\mu\nu})-J^{\mu}A_{\mu}.
\end{equation}
Using the Bianchi identity (18) and dropping the total derivative
terms, the expression (25) can also be written as follows:
\begin{equation}
{\cal L}=-\;\frac{1}{4\mu_{0}}
F_{\mu\nu}F^{\mu\nu}+\frac{1}{2\mu_{0}}\;a^{2}(\partial_{\sigma}F^{\rho\sigma})
(\partial^{\beta}F_{\rho\beta})-J^{\mu}A_{\mu}\;,
\end{equation}
where $a:=\hbar\sqrt{2\beta}$ . Equation (26) is the Lagrangian
density originally introduced by Podolsky [30-33], and $a$ is called
Podolsky's characteristic length [34-38]. The Euler-Lagrange
equation for the Lagrangian density (25) is [39,40]
\begin{equation}
\frac{\partial{\cal L}}{\partial A_{\lambda}
}-\partial_{\rho}\left(\frac{\partial{\cal
L}}{\partial(\partial_{\rho}A_{\lambda})}\right)+\partial_{\sigma}\partial_{\rho}\left(\frac{\partial{\cal
L}}{\partial(\partial_{\sigma}\partial_{\rho}A_{\lambda})}\right)=0.
\end{equation}
If we substitute (25) into (27), we will obtain the inhomogeneous
Maxwell's equations in the presence of a minimal observable
distance as follows:
\begin{equation}
\partial_{\rho}F^{\rho\lambda}+(\hbar\sqrt{2\beta})^{2}\;\Box\partial_{\rho}F^{\rho\lambda}=\mu_{0}J^{\lambda}.
\end{equation}
It should be mentioned that equations (28) have been previously
obtained from a different perspective by Kober [41]. Equations (18)
and (28) can be written in the vector form as follows:
\begin{eqnarray}
\nabla\cdot\textbf{E}+(\hbar\sqrt{2\beta})^{2}\;\Box(\nabla\cdot\textbf{E}) &=& \frac{\rho}{\varepsilon_{0}}\;,\\
\nabla\times\textbf{E} &=&-\;\frac{\partial \textbf{B}}{\partial
t}\;,\\
\nabla\times\textbf{B}+(\hbar\sqrt{2\beta})^{2}\;\Box(\nabla\times\textbf{B}-\frac{1}{c^{2}}\frac{\partial\textbf{E}}{\partial
t})&=&\mu_{0}\textbf{J}+\frac{1}{c^{2}}\frac{\partial\textbf{E}}{\partial
t},\\
\nabla\cdot\textbf{B}&=& 0.
\end{eqnarray}
The generalized Maxwell's equations (29)-(32) have been introduced
earlier by Tkachuk in Ref. [23] with a different approach. In the
limit $\hbar\sqrt{2\beta}\longrightarrow 0$, the generalized
inhomogeneous Maxwell's equations (29) and (31) become the usual
inhomogeneous Maxwell's equations.

\section{Free Space Solutions of the Generalized Inhomogeneous Maxwell's Equations}
In this section, we obtain the plane wave solutions of the
generalized inhomogeneous Maxwell's equations (28) in a
$3+1$-dimensional space-time.\\In free space
$(\rho=0,\textbf{J}=0)$, equations (28) can be written as
\begin{equation}
\partial_{\rho}F^{\rho\lambda}+(\hbar\sqrt{2\beta})^{2}\;\Box\partial_{\rho}F^{\rho\lambda}=0.
\end{equation}
In the Lorentz gauge $(\partial_{\rho}A^{\rho}=0)$ the field
equations (33) become
\begin{equation}
\Box A^{\lambda}+(\hbar\sqrt{2\beta})^{2}\;\Box \Box A^{\lambda}=0.
\end{equation}
Now, we try to find a plane wave solution of (34):
\begin{equation}
A^{\lambda}(x)=A \;e^{-\frac{i}{\hbar}p.x\;}\epsilon^{\lambda}(p),
\end{equation}
where $\epsilon^{\lambda}(p)$ is the polarization four-vector and
$A$ is a normalization constant. In the above equation
$p^{\mu}=(\frac{E}{c},\textbf{p})$ is the momentum four-vector. If
we substitute (35) in (34), we will obtain
\begin{equation}
p^{2}(1-\frac{a^{2}}{\hbar^{2}}\;p^{2})=0\;,
\end{equation}
where
$p^{2}=p_{\mu}p^{\mu}=(\frac{E}{c})^{2}-\textbf{p}^{2}$\;.\\Equation
(36) leads to the following energy-momentum relations
\begin{equation}
E^{2}=c^{2}\textbf{p}^{2},
\end{equation}
\begin{equation}
E^{2}=m_{_{eff}}^{2}\;c^{4}+c^{2}\textbf{p}^{2},
\end{equation}
where
\begin{equation}
m_{_{eff}}:=\frac{\hbar}{ac}\;.
\end{equation}
Equation (37) describes a massless vector particle whereas
equation (38) describes a massive vector particle with the
effective mass $m_{_{eff}}$.

\section{Upper Bound Estimation of the Minimal Length in Generalized Electrodynamics}
Substituting $\beta'=2\beta$ into (6), and remembering
$a=\hbar\sqrt{2\beta}$, we have
\begin{equation}
(\triangle X^{i})_{0}=\sqrt{(\frac{D+2}{2}) \;a^{2}\;[1+{\cal
O}\left(a^{2})\right]}\quad , \quad\forall i\in \{1,2, \cdots ,D\}.
\end{equation}
If we neglect terms of order $a^{4}$ and higher in (40), the
isotropic minimal length in a $3+1$-dimensional space-time becomes
\begin{equation}
(\triangle X^{i})_{0}\simeq{\frac{\sqrt{10}}{2}}\;a\quad ,
\quad\forall i\in \{1,2,3\}.
\end{equation}
Now we are ready to estimate the upper bounds on the isotropic
minimal length in generalized electrodynamics.
\subsection{Upper Bound on the Isotropic Minimal Length Based on the Anomalous Magnetic Moment of the Electron}
In a series of papers, Accioly and co-workers [27,29,34] have
estimated an upper bound on Podolsky's characteristic length $a$ by
computing the anomalous magnetic moment of the electron in the
framework of Podolsky's electrodynamics. This upper bound on $a$ is
[27,29,34]
\begin{equation}
a\leq\;4.7\times10^{-18}\;m.
\end{equation}
Inserting equation (42) into equations (39) and (41), we find
\begin{equation}
m_{_{eff}}\geq\; 41.8 \; \frac{GeV}{c^{2}}\;,
\end{equation}
\begin{equation}
(\triangle X^{i})_{0}\leq\; 7.4\times10^{-18}\;m.
\end{equation}
\subsection{Upper Bound on the Isotropic Minimal Length Based on the Ground State Energy of the Hydrogen Atom}
In Ref. [37], Cuzinatto and co-workers have studied the influence of
Podolsky's electrostatic potential on the ground state energy of the
hydrogen atom. In their study, the upper limit on $a$ is
\begin{equation}
a\leq\;5.56\times10^{-15}\;m.
\end{equation}
Inserting equation (45) into equations (39) and (41), we find
\begin{equation}
m_{_{eff}}\geq\; 35.51 \; \frac{MeV}{c^{2}}\;,
\end{equation}
\begin{equation}
(\triangle X^{i})_{0}\leq\; 8.79\times10^{-15}\;m.
\end{equation}
It should be noted that the upper bound (47) is about three
orders of magnitude larger than the upper bound (44), i.e.,
\begin{equation}
(\triangle
X^{i})_{0}{_{\;Ground\;State\;Energy\;of\;the\;Hydrogen\;
Atom}}\,\sim 10^{3}\;(\triangle
X^{i})_{0}{_{\;Anomalous\;Magnetic\;
Moment\;of\;the\;Electron}}\;,
\end{equation}
while the lower bound (46) is about three orders of magnitude
smaller than the lower bound (43), i.e.,
\begin{equation}
m_{_{eff-\;Ground\;State\;Energy\;of\;the\;Hydrogen\; Atom}}\,\sim
\,10^{-3}\;m_{_{eff-\;Anomalous\;Magnetic\;
Moment\;of\;the\;Electron}}\;.
\end{equation}

\section{Relationship between Nonlocal Electrodynamics and Electrodynamics in the Presence of a Minimal Length}
In a series of papers, Smailagic and Spallucci [42-44] have
introduced an approach to formulate non-commutative quantum field
theory. Using Smailagic-Spallucci approach, Gaete and Spallucci
introduced a nonlocal Lagrangian density for the vector field
$A^{\mu}$ with an external source $J^{\mu}$ as follows:
\begin{equation}
{\cal L}=-\;\frac{1}{4\mu_{0}}
F_{\mu\nu}\exp{(\theta\;\Box)}\;F^{\mu\nu}-J^{\mu}A_{\mu}\;,
\end{equation}
where $\theta$ is a constant parameter with dimensions of
${(length)}^2$ [45]. We assume that the function
$\exp{(\theta\;\Box)}$ in (50) can be expanded in a power series as
follows:
\begin{equation}
\exp{(\theta\;\Box)}=\sum_{l=0}^{\infty}
\frac{\theta^l}{l\;!}\;\Box\;^{l},
\end{equation}
where $\Box\;^{l}$ denotes the $\Box$ operator applied $l$ times [46].\\
If we insert (51) into (50), we will obtain the following Lagrangian
density
\begin{equation}
{\cal L}=-\;\frac{1}{4\mu_{0}}
F_{\mu\nu}F^{\mu\nu}-\frac{1}{4\mu_{0}}\;\theta F_{\mu\nu}\;\Box
F^{\mu\nu}-\frac{1}{4\mu_{0}}\sum_{l=2}^{\infty}\frac{\theta\;^l}{l\;!}\;F_{\mu\nu}\;\Box\;^{l}F^{\mu\nu}-J^{\mu}A_{\mu}\;.
\end{equation}
After neglecting terms of order $\theta^{2}$ and higher in (52) we
obtain
\begin{equation}
{\cal L}=-\;\frac{1}{4\mu_{0}}
F_{\mu\nu}F^{\mu\nu}-\frac{1}{4\mu_{0}}\;\theta F_{\mu\nu}\;\Box
F^{\mu\nu}-J^{\mu}A_{\mu}\;.
\end{equation}
A comparison between equations (22) and (53) shows that there is
an equivalence between the Gaete-Spallucci electrodynamics to
first order in $\theta$ and the Lee-Wick electrodynamics (or
electrodynamics in the presence of a minimal length).
The
relationship between the non-commutative parameter $\theta$ in
(53) and $a=\hbar\sqrt{2\beta}$ in (22) is
\begin{equation}
a=\sqrt{\theta}.
\end{equation}
Inserting equation (54) into equations (39) and (41), we find
\begin{equation}
m_{_{eff}}=\frac{\hbar}{\sqrt{\theta}c}\;,
\end{equation}
\begin{equation}
(\triangle X^{i})_{0}\simeq{\frac{\sqrt{10\;\theta}}{2}}\quad ,
\quad\forall i\in \{1,2,3\}.
\end{equation}
Using (45) in (54), we obtain the following upper bound for the
non-commutative parameter $\theta$:
\begin{equation}
\theta_{_{Ground\;State\;Energy\;of\;the\;Hydrogen\; Atom}}\leq \;
3.09\times 10^{-29}\; m^{2}.
\end{equation}
The above upper bound on the non-commutative parameter $\theta$,
i.e., $3.09\times 10^{-29}\; m^{2}$ is near to the neutron-proton
scattering cross section $(10^{-25}\;cm^{2})$ [47]. It is necessary
to note that the electrodynamics in the presence of a minimal
observable distance is only correct to the first order in the
deformation parameter $\beta$, while the Gaete-Spallucci
electrodynamics is valid to all orders in the non-commutative
parameter $\theta$.

\section{Conclusions}
Heisenberg believed that every theory of elementary particles should
contain a minimal observable distance of the magnitude $\ell_{0}\sim
10^{-13}\,cm$ [47-50]. He hoped that the introduction of a minimal
length would eliminate divergences that appear in quantum
electrodynamics. Today's we know that every theory of quantum
gravity predicts the existence of a minimal measurable length which
leads to a GUP. An immediate consequence of the GUP is a
generalization of position and derivative operators according to
equations (19) and (20) for $\beta'=2\beta$. It was shown that the
Lagrangian formulation of electrodynamics with an external source in
the presence of a minimal measurable length leads to the
inhomogeneous fourth-order field equations. We demonstrated the
similarity between electrodynamics in the presence of a minimal
length and Lee-Wick electrodynamics. We have shown that the free
space solutions of the inhomogeneous Maxwell's equations in the
presence of a minimal length describe two particles, a massless
vector particle and a massive vector particle with the effective
mass $m_{_{eff}}=\frac{\hbar}{ac}$. Now, let us compare the upper
bounds on the isotropic minimal length in this paper with the
results of Refs. [47-51]. The upper limit on the isotropic minimal
length in equation (44) is near to the electroweak length scale
$(\ell_{electroweak}\sim 10^{-18}\, m)$ [51], while the upper limit
(47) is near to the minimal observable distance which was proposed
by Heisenberg $(\ell_{0}\sim 10^{-13}\,cm)$ [47-50]. It is
interesting to note that the lower bound on the effective mass
$m_{_{eff}}$ in equation (43), i.e., $41.8\;\frac{GeV}{c^{2}}$ is of
the same order of magnitude as the mass of the $W^{\pm}$ and $Z^{0}$
vector bosons $(M_{_{W}}=80.425\pm0.038\;\frac{GeV}{c^{2}}\; , \;
M_{_{Z}}=91.1876\pm0.0021\;\frac{GeV}{c^{2}})$ [52]. Finally, we
have investigated the relationship between the Gaete-Spallucci
nonlocal electrodynamics and electrodynamics with a minimal length.

\section*{Note added}
After this work was completed, we became aware of an interesting
article by Maziashvili and Megrelidze [53], in which the authors
study the electromagnetic field in the presence of a momentum
cutoff. For their discussion they use the following modified
Heisenberg algebra
\begin{eqnarray}
\left[X^{i},P^{j}\right] &=&i\hbar\;(\frac{2\beta \textbf{P}^{2}}{\sqrt{1+4\beta \textbf{P}^{2}}-1}\;\delta^{ij}+2\beta P^{i}P^{j}), \\
\left[X^{i},X^{j}\right] &=& 0, \\
\left[P^{i},P^{j}\right] &=& 0,
\end{eqnarray}
where $i,j=1,2,3$ and $\beta$ is a deformation parameter [54]. In
our work we have formulated electrodynamics in the framework of
Quesne-Tkachuk algebra which is a Lorentz-covariant deformed algebra
whereas the authors of Ref. [53] have studied electrodynamics in the
framework of (58)-(60) algebra which is not a Lorentz-covariant
algebra.

\section*{Acknowledgments}
We would like to thank the referees for their useful comments.


\begin{thebibliography}{11}
\bibitem{P1}
M. Sprenger, P. Nicolini and M. Bleicher, Eur. J. Phys.
\textbf{33}, 853 (2012).
\bibitem{P2}
S. Hossenfelder, arXiv:1203.6191v1.
\bibitem{P3}
C. Castro, J. Phys. A: Math. Gen. \textbf{39}, 14205 (2006).
\bibitem{P4}
Y. Ko, S. Lee and S. Nam, Int. J. Theor. Phys. \textbf{49}, 1384
(2010).
\bibitem{P5}
S. Hossenfelder, Phys. Rev. D \textbf{70}, 105003 (2004).
\bibitem{P6}
M. S. Berger and M. Maziashvili, Phys. Rev. D \textbf{84}, 044043
(2011).
\bibitem{P7}
M. Kober, Int. J. Mod. Phys. A \textbf{26}, 4251 (2011).
\bibitem{P8}
S. Das and E. C. Vagenas, Phys. Rev. Lett. \textbf{101}, 221301
(2008).
\bibitem{P9}
A. F. Ali, S. Das and E. C. Vagenas, Phys. Lett. B \textbf{678},
497 (2009).
\bibitem{P10}
S. Das, E. C. Vagenas and A. F. Ali, Phys. Lett. B \textbf{690},
407 (2010).
\bibitem{P11}
A. F. Ali, S. Das and E. C. Vagenas, Phys. Rev. D \textbf{84},
044013 (2011).
\bibitem{P12}
C. Quesne and V. M. Tkachuk, Phys. Rev. A \textbf{81}, 012106
(2010).
\bibitem{P13}
S. K. Moayedi, M. R. Setare and H. Moayeri, Int. J. Theor. Phys.
\textbf{49}, 2080 (2010).
\bibitem{P14}
S. K. Moayedi, M. R. Setare and H. Moayeri, Europhys. Lett.
\textbf{98}, 50001 (2012).
\bibitem{P15}
S. Hossenfelder, M. Bleicher, S. Hofmann, J. Ruppert, S. Scherer and
H. Stocker, Phys. Lett. B \textbf{575}, 85 (2003).
\bibitem{P16}
M. R. Setare, Phys. Rev. D \textbf{70}, 087501 (2004).
\bibitem{P17}
S. Basilakos, S. Das and E. C. Vagenas, JCAP \textbf{09}, 027
(2010).
\bibitem{P18}
W. Chemissany, S. Das, A. F. Ali and E. C. Vagenas, JCAP
\textbf{12}, 017 (2011).
\bibitem{P19}
H. S. Snyder, Phys. Rev. \textbf{72}, 68 (1947).
\bibitem{P20}
C. Quesne and V. M. Tkachuk, J. Phys. A: Math. Gen. \textbf{39},
10909 (2006).
\bibitem{P21}
C. Quesne and V. M. Tkachuk, Czech. J. Phys.  \textbf{56}, 1269
(2006).
\bibitem{P22}
H. S. Snyder, Phys. Rev. \textbf{71}, 38 (1947).
\bibitem{P23}
V. M. Tkachuk, J. Phys. Stud. \textbf{11}, 41 (2007).
\bibitem{P24}
D. Griffiths, Introduction to Elementary Particles (Wiley, New York,
1987).
\bibitem{P25}
T. Lee and G. Wick, Nucl. Phys. B \textbf{9}, 209 (1969).
\bibitem{P26}
T. Lee and G. Wick, Phys. Rev. D \textbf{2}, 1033 (1970).
\bibitem{P27}
A. Accioly and E. Scatena, Mod. Phys. Lett. A \textbf{25}, 269
(2010).
\bibitem{P28}
A. Accioly, P. Gaete, J. Helayel-Neto, E. Scatena and R. Turcati,
arXiv:1012.1045v2.
\bibitem{P29}
A. Accioly, P. Gaete, J. Helayel-Neto, E. Scatena and R. Turcati,
Mod. Phys. Lett. A \textbf{26}, 1985 (2011).
\bibitem{P30}
B. Podolsky, Phys. Rev. \textbf{62}, 68 (1942) .
\bibitem{P31}
B. Podolsky and C. Kikuchi, Phys. Rev.  \textbf{65}, 228 (1944).
\bibitem{P32}
B. Podolsky and C. Kikuchi, Phys. Rev.  \textbf{67}, 184 (1945).
\bibitem{P33}
B. Podolsky and P. Schwed, Rev. Mod. Phys. \textbf{20}, 40 (1948).
\bibitem{P34}
A. Accioly and H. Mukai, Nuovo Cimento B \textbf{112}, 1061
(1997).
\bibitem{P35}
A. Accioly and H. Mukai, Braz. J. Phys. \textbf{28}, 35 (1998).
\bibitem{P36}
R. R. Cuzinatto, C. A. M. de Melo and P. J. Pompeia, Ann. Phys.
\textbf{322}, 1211 (2007).
\bibitem{P37}
R. R. Cuzinatto, C. A. M. de Melo, L. G. Medeiros and P. J.
Pompeia, Int. J. Mod. Phys. A \textbf{26}, 3641 (2011).
\bibitem{P38}
M. V. S. Fonseca and A. V. Paredes, Braz. J. Phys. \textbf{40}, 319
(2010).
\bibitem{P39}
J. Magueijo, Phys. Rev. D \textbf{73}, 124020 (2006).
\bibitem{P40}
C. M. Reyes, Phys. Rev. D \textbf{80}, 105008 (2009).
\bibitem{P41}
M. Kober, Phys. Rev. D \textbf{82}, 085017 (2010).
\bibitem{P42}
A. Smailagic and E. Spallucci, J. Phys. A: Math. Gen. \textbf{36},
L517 (2003).
\bibitem{P43}
A. Smailagic and E. Spallucci, J. Phys. A: Math. Gen. \textbf{36},
L467 (2003).
\bibitem{P44}
A. Smailagic and E. Spallucci, J. Phys. A: Math. Gen. \textbf{37},
7169 (2004).
\bibitem{P45}
P. Gaete and E. Spallucci, J. Phys. A: Math. Theor. \textbf{45},
065401 (2012).
\bibitem{P46}
J. E. Lidsey, Int. J. Mod. Phys. D \textbf{17}, 577 (2008).
\bibitem{P47}
T. G. Pavlopoulos, Phys. Rev. \textbf{159}, 1106 (1967).
\bibitem{P48}
W. Heisenberg, Z. Naturforschr, \textbf{5A}, 251 (1950).
\bibitem{P49}
W. Heisenberg, Ann. Physik \textbf{32}, 20 (1938).
\bibitem{P50}
T. G. Pavlopoulos, Phys. Lett. B \textbf{625}, 13 (2005).
\bibitem{P51}
B. Zwiebach, A First Course in String Theory, Second Edition
(Cambridge University Press, 2009).
\bibitem{P52}
W. N. Cottingham and D. A. Greenwood, An Introduction to the
Standard Model of Particle Physics, Second Edition (Cambridge
University Press, 2007).
\bibitem{P53}
M. Maziashvili and L. Megrelidze, arXiv:1212.0958v1.
\bibitem{P54}
A. Kempf and G. Mangano, Phys. Rev. D \textbf{55}, 7909 (1997).


\end{thebibliography}
\end{document}